\documentclass[aps,pra]{revtex4}
\begin{document}

\title{Comment on ' Eavesdropping on the ping-pong quantum
communication protocol'
\thanks{E-mail: Zhangzj@wipm.ac.cn}}

\author{
Z. J. Zhang \\
{\footnotesize  Wuhan Institute of Physics and Mathematics, The
Chinese Academy of Sciences, Wuhan 430071, China} }

\date{\today}

\maketitle Recently W\'{o}jcik has analyzed the security of the
'ping-pong' quantum communication protocol[1] in the case of
considerable quantum channel losses and accordingly an
undetectable eavesdropping scheme and possible improvements on the
'ping-pong' protocol are proposed [2]. This is true. Moreover,
according to the analysis on the mutual information $I$ as a
function of the transmission efficiency $\eta$ of the quantum
channel, W\'{o}jcik concludes that the 'ping-pong' protocol is not
secure for $\eta < 60\%$. The aim of this Comment is to point out
that the conclusion is not reliable.

From the figure 4 in W\'{o}jcik's paper, one can see that when
$\eta=0$, both $I_{AB}$ (the mutual information between two
legitimate users, say, Alice as the message sender and Bob as the
message receiver) and $I_{AE}$ (the mutual information between
Alice and the eavesdropper Eve) are nonzero and $I_{AE}$ is larger
than $I_{AB}$. This means that even if no message is transmitted
by Alice,  both Eve and Bob still get some information from Alice,
and Eve gets more information than Bob. This is unimaginable.

In W\'{o}jcik's paper, the theoretical $I_{AE} (I_{AB})$ is always
0.311 (0.189 after the symmetrization).  In W\'{o}jcik's scheme (a
realistic scheme, where the number of the transmitted bits should
be finite), due to Eve's attacks, Eve's (Bob's) final bits can not
be predicted deterministically. Alternatively, for a certain batch
of Alice's bits, provided that Eve's attacks are given, different
batches of bits can be obtained by Eve (Bob) with different
possibilities in theory. Since the batches of Eve's (Bob's) bits
are not unique in theory, and generally speaking, different
batches of bits should lead to different mutual information, hence
in the scheme the mutual information between Alice and Eve (Bob)
should not be unique in theory too. This can be easily verified by
a simple example as follows. Let 'u' ('s') be Eve's attack without
(with) the symmetry operation. Assume Alice's bits are '100110'
and Eve's attacks are 'susuus'. According to W\'{o}jcik's scheme,
taking advantage of the following conditional probability
distributions:
\begin{eqnarray}
p^u_{000}=1, \,\, p^u_{001}=p^u_{010}=p^u_{011}=0, \,\,\,\,\,\,
p^u_{100}=p^u_{101}=p^u_{110}=p^u_{111}=1/4; \nonumber \\
p^s_{000}=p^s_{001}=p^s_{010}=p^s_{011}=1/4, \,\,\,\,\,\,\,
p^s_{111}=1, \,\, p^s_{100}=p^s_{101}=p^s_{110}=0;
\end{eqnarray}
in theory Eve might get any one of the following batches:
'100110', '100111', '100100', '100101', '100010', '100011',
'100000', '100001','101100', '101101', '101110', '101111',
'101000', '101001','101010' and '101011'.  The possibility, the
QBER and the mutual information for each batch of bits are shown
in Table 1.

\newpage

\noindent Table 1  Possibility, QBER, mutual information for Eve's
possible batches of bits.
\begin{center}
\begin{tabular}{ccccccc} \hline
Eve's bits & possibility & ${\rm QBER}$ &&  && $I_{AE}$ \\ \hline
'100110'   & 1/16        &  0  &&   &&    1     \\
'100111'   & 1/16        & 1/6 &&    &&   0.459  \\
'100100'   & 1/16        & 1/6 &&    &&   0.459  \\
'100101'   & 1/16        & 1/3 &&    &&   0.082  \\
'100010'   & 1/16        & 1/6 &&    &&   0.459  \\
'100011'   & 1/16        & 1/3 &&    &&   0.082  \\
'100000'   & 1/16        & 1/3 &&    &&   0.134  \\
'100001'   & 1/16        & 1/3 &&    &&   0.093  \\
'101100'   & 1/16        & 1/3 &&    &&   0.082  \\
'101101'   & 1/16        & 1/2 &&    &&     0    \\
'101110'   & 1/16        & 1/6 &&    &&   0.459  \\
'101111'   & 1/16        & 1/3 &&    &&   0.093  \\
'101000'   & 1/16        & 1/2 &&    &&     0    \\
'101001'   & 1/16        & 2/3 &&    &&   0.082  \\
'101010'   & 1/16        & 1/3 &&    &&  0.082   \\
'101011'   & 1/16        & 1/2 &&    &&    0     \\ \hline
\end{tabular} \\
Alice's bits  are '100110'. Eve's attacks are 'susuus'. \\ The
transmission efficiency $\eta$ is assumed to be not greater than
50\%.
\end{center}

To summarize, there must be mistakes in W\'{o}jcik's calculations
of the mutual information. Concluding, the security estimation
based on the wrong mutual information is not reliable anymore.

This work is funded by the National Science Foundation of China
(Grant No.10304022).\\

\noindent [1] K. Bostr\"{o}m and T. Felbinger, Phys. Rev. Lett.
{\bf 89}, 187902(2002).

\noindent [2] Antoni W\'{o}jcik, Phys. Rev. Lett., {\bf 90},
157901(2003).

\end{document}